\documentclass[twocolum]{IEEEtran}
\usepackage{amsmath, graphics,amssymb,epsfig,subfigure}
\usepackage{etoolbox}
\usepackage{cite}
\usepackage{url}
\usepackage{algorithm}
\usepackage{color,soul}
\usepackage{cases}
\usepackage{epstopdf}
\usepackage{graphicx,amsmath}
\makeatletter
\patchcmd{\@makecaption}
  {\scshape}
  {}
  {}
  {}
\makeatother

\begin{document}
\title{Deep Learning Framework for Wireless Systems: Applications to Optical Wireless Communications}
%%%%%%%%%%%%%%%%%%%%%%%%%%%%%%%%%%%%%%%%%%%%%%%%%%%%%%%%%%%%%%%%%%%%%%%%%%%%%%%%%%%%%%%%%%%%%%
\author{\IEEEauthorblockN{Hoon Lee, \textit{Member}, \textit{IEEE}, Sang Hyun Lee, \textit{Member}, \textit{IEEE},\\
                          Tony Q. S. Quek, \textit{Fellow}, \textit{IEEE}, Inkyu Lee, \textit{Fellow}, \textit{IEEE}}\\
\thanks{This work was supported by the National Research Foundation through the Ministry of Science, ICT and Future Planning (MSIP), South Korean Government, under Grant 2017R1A2B3012316. The work of S. H. Lee was supported in part by Institute for Information \& communications Technology Promotion (IITP) grant funded by the Korea government (MSIT) (2016-0-00208, High Accurate Positioning Enabled MIMO Transmission and Network Technologies for Next 5G-V2X (vehicle-to-everything) Services). The work of T. Q. S. Quek was supported by the SUTD-ZJU Research Collaboration under Grant SUTD-ZJU/RES/05/2016.

H. Lee and T. Q. S. Quek are with the Information Systems Technology and Design Pillar, Singapore University of Technology and Design, Singapore 487372 (e-mail: \{hoon$\_$lee, tonyquek\}@sutd.edu.sg). S. H. Lee and I. Lee are with the School of Electrical Engineering, Korea University, Seoul 02841, South Korea (e-mail: \{sanghyunlee, inkyu\}@korea.ac.kr).

\copyright 2018 IEEE. Personal use of this material is permitted. Permission from IEEE must be obtained for all other uses, in any current or future media, including reprinting/republishing this material for advertising or promotional purposes, creating new collective works, for resale or redistribution to servers or lists, or reuse of any copyrighted component of this work in other works.}
}\maketitle \thispagestyle{empty}
%%%%%%%%%%%%%%%%%%%%%%%%%%%%%%%%%%%%%%%%%%%%%%%%%%%%%%%%%%%%%%%%%%%%%%%%%%%%%%%%%%%%%%%%%%%%%%%%%%%%%%%%Abstract
%\vspace{-20mm}
\begin{abstract}
Optical wireless communication (OWC) is a promising technology for future wireless communications owing to its potentials for cost-effective network deployment and high data rate. There are several implementation issues in the OWC which have not been encountered in radio frequency wireless communications. First, practical OWC transmitters need an illumination control on color, intensity, and luminance, etc., which poses complicated modulation design challenges. Furthermore, signal-dependent properties of optical channels raise non-trivial challenges both in modulation and demodulation of the optical signals. To tackle such difficulties, deep learning (DL) technologies can be applied for optical wireless transceiver design. This article addresses recent efforts on DL-based OWC system designs. A DL framework for emerging image sensor communication is proposed and its feasibility is verified by simulation. Finally, technical challenges and implementation issues for the DL-based optical wireless technology are discussed.
\end{abstract}

\section{Introduction}
Optical wireless communication (OWC), which exploits terahertz spectra, has been regarded as a promising solution for enabling much higher data rate in the fifth generation (5G) communication systems~\cite{Khalighi:14}. In OWC, solid-state optical sources such as light-emitting diodes (LEDs) are used to convey information to receivers equipped with photodiodes (PDs) and serve as lighting sources at the same time by switching optical pulses at a very high rate such that human eyes cannot perceive. Compared to radio frequency (RF) communication, a more cost-effective deployment is possible for OWC by utilizing existing lighting infrastructures and leveraging spectrum which does not require authorized access.

In practical implementation of optical wireless systems, the average intensity of LEDs is controlled for achieving energy savings and safety \cite{Khalighi:14}. In particular, the average intensity of optical pulses in visible light communication (VLC) applications is closely related to color and luminance of LEDs which are essential features for user requirement satisfaction. Thus, the optical signal modulation and demodulation strategies should be designed to fulfill arbitrary intensity constraints. Various optical modulation techniques have been developed with the objective of improving spectral efficiency \cite{SZhao:17} and maximizing the minimum distance among constellation points \cite{XLiang:17,Ostergard:10}. However, the optimal design with respect to end-to-end error rate performance under generic lighting constraints still remains unaddressed. Another critical issue is the optical shot noise induced by random nature of the photon emission of LEDs. The statistics of the optical shot noise depends on the transmitted LED intensity, and thus the application of transceiver design methods intended for RF communication results in a performance loss.

To tackle such non-trivial challenges, this article presents deep learning (DL) techniques that identify an efficient optical transceiver pair. In particular, an unsupervised learning framework based on autoencoder (AE) \cite{YBengio:09} is investigated to design an optical wireless system. The AE techniques have been recently adopted to RF communication designs \cite{OShea:17}. However, it is not straightforward to bring the machine learning structure in \cite{OShea:17} into the OWC design since the effect of lighting constraints has not been properly studied. Therefore, additional processing is required to control behaviors of neural networks (NNs) to construct OWC systems. Since most state-of-the-art DL techniques have focused on unconstrained problems in classification and generative model applications, it is highly challenging to introduce complicated constraints into the NNs in general.

This article provides an overview of recent DL approaches for various optical wireless setups such as multi-colored systems and on-off keying (OOK) based OWC. Subsequently, a convolutional AE (C-AE) structure is proposed for image sensor communication (ISC) where the information is conveyed by spatially separated LED arrays and a receiver is implemented with an optical image sensor. Finally, concluding remarks and implementation challenges for DL-based communication systems are addressed.

\section{Deep Learning Framework for OWC Systems}

\subsection{AE Basics}

\begin{figure*}[!htp]
\begin{center}
\includegraphics[width=5.0in]{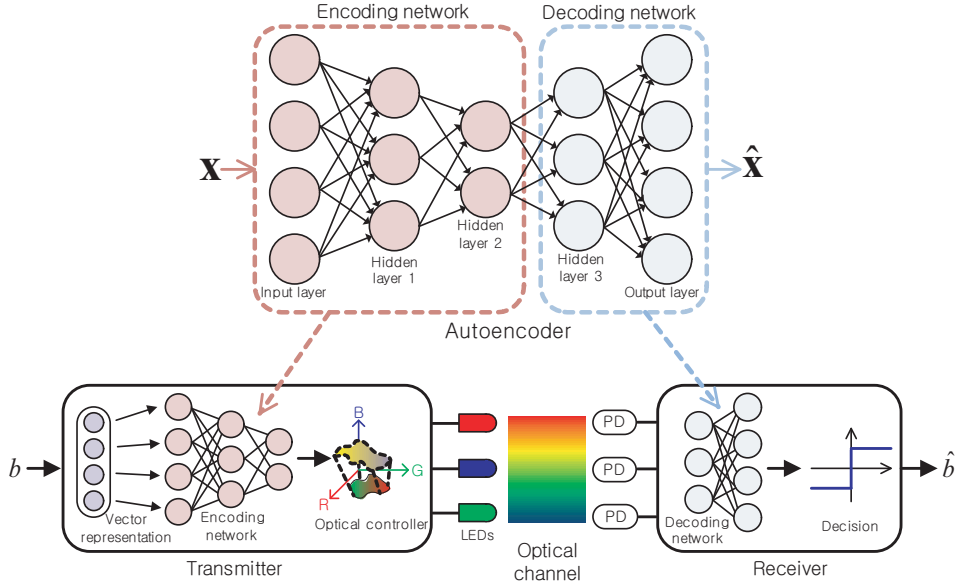}
\end{center}
\caption{AE construction for OWC design.}
\label{fig:fig1}
\end{figure*}

Figure \ref{fig:fig1} shows an AE which consists of an input layer, multiple hidden layers, and an output layer. The objective of the AE is to find efficient encoding and decoding rules for a given training set without any prior knowledge. The AE framework focuses on an accurate reconstruction of input $\mathbf{x}$ by its output $\hat{\mathbf{x}}$ obtained from successive NN computations. To learn the encoding and decoding rules effectively, the dimension of hidden layers of a typical AE first decreases and subsequently increases after a certain layer (e.g., hidden layer 2 in Fig. \ref{fig:fig1}). Thus, the AE can be regarded as a cascade of two consecutive NNs: an encoding network and a decoding network. The output of the encoding network can be interpreted as a codeword, while the decoding network produces a reconstruction from the codeword such that the output becomes similar to the input.

Each hidden layer performs a linear operation on an input with a trained set of weight matrices and bias vectors and applies a nonlinear activation at the result of the linear operation to yield the final output. The reconstruction is then attained at the output layer via a similar computation process. The activation is an important feature in DL which introduces non-linearity to NNs so that complicated input-output relationships can be effectively learned. Popular candidates include rectified linear unit (ReLU), sigmoid, and softmax \cite{LeCun:15}.

During the training step of the AE, the weight matrices and the bias vectors of the hidden layers and the output layer are trained to minimize a cost function of the AE, which assesses the affinity between the input and the output for successful reconstruction processes. For continuous-valued inputs, the mean square error is typically adopted for the cost function, whereas the cross-entropy works well in classification applications with binary inputs. The minimization of the AE cost function is, in general, a non-convex optimization problem where no closed-form solution is available. To train an NN, most state-of-the-art DL techniques employ a stochastic gradient descent (SGD) algorithm, which is a variant of the gradient descent methods \cite{LeCun:15}. Once the AE is trained, its performance is evaluated over a test set whose elements are not seen during the training step.

\subsection{Applications to OWC Systems}
Figure \ref{fig:fig1} illustrates a generic AE framework for OWC systems. The AE-based optical transceiver is implemented with the encoding and the decoding networks which have been trained in advance. A message index $b$ in the message set $\{1,\cdots,M\}$ is first mapped to a vector representation to be processed by multi-dimensional hidden layers. The representation vector is passed into the encoding network at the transmitter, and an optical controller, in turn, refines the output of the encoding network to produce feasible optical signals $\mathbf{s}_{b}$ of length $N$ for each message $b$. The output dimension of the transmitter corresponds to the number of LEDs or the symbol duration. The optical controller can be realized by either deterministic computation or additional NN that is trained along with the encoding and the decoding networks.

The optical channel can be characterized by two different types of noise sources: ambient noise and shot noise. The ambient noise is independent of the transmitted signal and is assumed as a Gaussian random variable with zero mean and variance $\sigma^{2}$. On the other hand, the shot noise is induced by the random nature of photon emission of LEDs and its variance is proportional to the channel input. Thus, for the channel input $s$, the shot noise variance is modeled as $\psi^{2}\sigma^{2}s$, where $\psi^{2}$ stands for the shot noise scaling factor. Finally, the output $\hat{b}$ obtained from the decoding network at the receiver is an estimate of the transmitted message $b$.

In AE techniques which have been recently presented for RF communication system designs \cite{OShea:17}, transmit power constraint of practical RF hardware is fulfilled by a simple deterministic normalization in the two-dimensional (2D) RF signal space \cite{OShea:17}. In contrast, lighting constraints in OWC are normally interpreted by polyhedrons in a multi-dimensional space, which forms a more complicated optical signal space compared to the RF communication. Furthermore, some optical systems such as an OOK-based OWC system are subject to non-convex optical constraints. Therefore, handling the behavior of the encoding network based on simple computations in \cite{OShea:17} is not straightforward for the OWC systems. As a result, the design of the optical controller is a key challenge for the AE-based OWC optimization.

\section{Deep Learning Based OWC Design}
In this section, recent technical progresses of the AE methods in OWC transceiver design problems are presented.

\subsection{Multi-color Modulation}
A VLC system is one of a multi-colored OWC and consists of multi-color LEDs with $N$ color chips which send messages to a receiver with $N$ corresponding PDs. Different color filters are utilized at the receiver to separate optical signals according to the corresponding color. Each message is modulated to an optical constellation point in the $N$-dimensional color signal space. An optical space is specified by constraints on the non-negativity and the peak intensity for each color dimension. Also, the average intensity of the optical modulation signal should meet the dimming target which is associated with color and intensity requirements of users.

The AE framework has been applied to the multi-colored VLC systems in \cite{HLee:18} for identifying a reliable message recovery technique. To add the dimming support, a deterministic post-processing computation is carried out at the end of the encoding network. A convex optimization formulation is employed to project the output vector of the encoding network onto the feasible optical space. Based on a closed-form projection solution, a low-complexity implementation is possible for training numerous samples.

The optical channel is adopted with a stochastic noise layer where a randomly generated Gaussian noise vector is added to the transmitted signal. Then, the decoding network receives the noisy signal as an input and performs a classification task with the softmax output activation. This produces a probability vector where each element characterizes the probability of the corresponding message being transmitted. The cost function is set to the categorial cross-entropy function between the input message and the output probability vector for the classification task of the transmitted message. Thus, the training set of the AE consists of a large number of messages and optical noise vectors. In the training step, the AE learns dimming features as well as statistical properties of the optical channels by itself to minimize the classification error. It has been noticed that in the signal-dependent shot noise channels, the AE method performs better than classical minimum distance maximizing approaches \cite{XLiang:17} in terms of the average symbol error rate (SER) performance and effectively mitigates the effects of inter-color interference induced by the imperfection of received color filters.

\subsection{OOK Modulation}
\begin{figure*}[!htp]
\begin{center}
\includegraphics[width=5.0in]{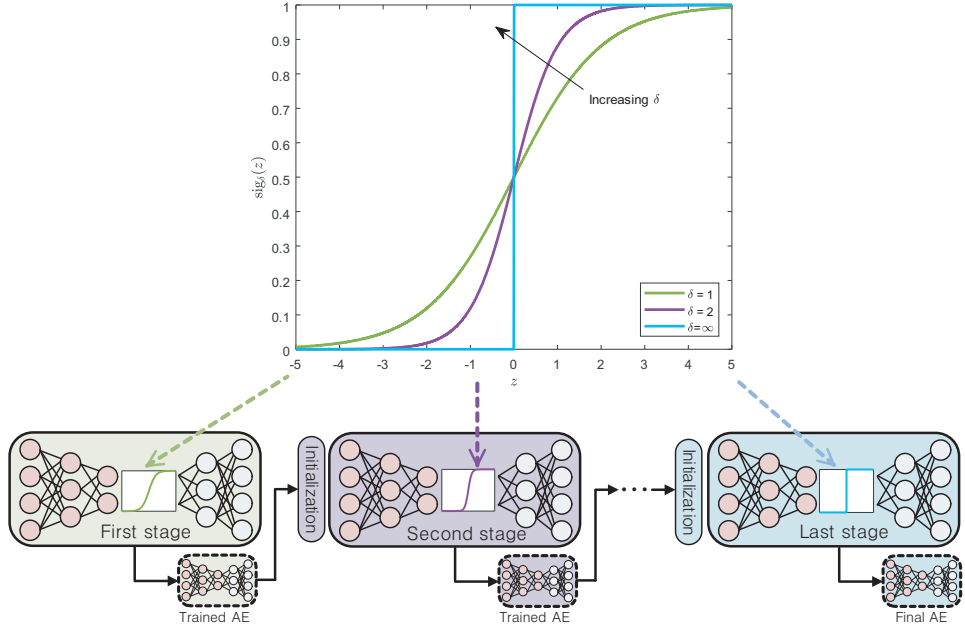}
\end{center}
\caption{Concept of soft binarization approaches and multi-stage training strategy.}
\label{fig:fig2}
\end{figure*}

In OOK-based OWC, each LED either turns on or off to convey a binary message. Thus, a message is encoded by a binary optical signal whose average intensity is controlled by adjusting the number of ones in the binary vector. This requires a computationally demanding search for designing constant weight codes (CWCs), which consist of binary codewords with identical Hamming weight. Although the mathematical properties of the CWCs have been intensively studied \cite{Ostergard:10}, identifying the optimal encoding and decoding rules for the CWC still remains open in general configurations.

In \cite{HLee:18b}, a binary AE training approach has been presented for an OOK-based OWC where a message is conveyed through temporal intensity change of a single LED. Thus, the output dimension $N$ of the transmitter indicates the symbol duration, which is the length of the binary codeword. To restrict the number of ones in the binary codeword $\mathbf{s}_{b}=[s_{b,1},\cdots,s_{b,N}]^{T}$ of each message $b$, the penalty term $\lambda(\sum_{j=1}^{N}s_{b,j}-d)^{2}$ is augmented to the AE cost function, where a positive number $\lambda$ represents a tradeoff parameter controlling the portion of the penalty term in the cost function and $d$ stands for the target average intensity.

Since the binary constraint $s_{b,j}\in\{0,1\}$ is non-convex, deterministic operations including linear projections in \cite{HLee:18} are no longer applicable to the OOK systems. A naive approach for generating binary outputs would be to employ a hard binary activation such as a unit step function. However, the gradient of the hard binarization function is zero for all input range and must incur a well-known \textit{vanishing gradient} problem \cite{LeCun:15}, which is a notorious issue in training deep NNs handling discrete variables. Hence, the weights and the biases of the AE do not get updated with the SGD algorithm and are typically stuck with poor performance.

To overcome this difficulty, a soft binarization technique has been adopted in \cite{HLee:18b} which gradually anneals a continuous-valued latent vector into an OOK signal during the training step. At the end of the encoding layer, a parameterized sigmoid function $\text{sig}_{\delta}(z)=\frac{1}{1+\exp(-\delta z)}$ is utilized as the activation function, where a positive number $\delta$ is related to the tangent of the sigmoid function. As illustrated in Fig. \ref{fig:fig2}, the parameterized sigmoid function approaches the hard binary activation as $\delta$ becomes larger. Thus, the hardness of the AE is controlled by adjusting the parameter~$\delta$.

For a moderate regime of $\delta$, the parameterized sigmoid function has a non-zero gradient, implying that the AE can be efficiently trained via the SGD algorithm. To avoid the vanishing gradient problem with a large value of $\delta$, a multi-stage training strategy has been introduced in \cite{HLee:18b} which sequentially trains the AE with a different $\delta$ at each stage. The parameter $\delta$ is gradually incremented at each stage so that the training performance converges to an effective point without the vanishing gradient issue.

Figure \ref{fig:fig2} depicts a multi-stage training strategy. At each stage, weights and biases of the AE are trained with $\delta$ fixed using the SGD algorithm until convergence. Upon the training completion, the value of $\delta$ is incremented for training at the next stage. Thus, the SGD algorithm at the current stage {\em warm-starts} from the AE trained at the previous stage. It can be viewed as a cascaded fine-tuning strategy of an AE. Finally, the value of $\delta$ at the last stage becomes a sufficiently large number such that a binary output is produced. The DL approach for the OOK-based OWC outperforms traditional minimum Hamming distance maximization designs \cite{Ostergard:10} in terms of the average SER over the shot noise channels.

\section{Deep Learning Framework for Image Sensor Communication}
The DL approaches reviewed in the previous section were developed for an optical receiver with a single PD which forms parallel single-input single-output channels. This section proposes the extension of such results for an ISC system where a receiver is realized by a image sensor which consists of multiple PDs. Thus, the ISC can be regarded as a multiple-input multiple-output optical wireless system.
The use of the image sensor as an OWC receiver has been intensively studied in recent years, and its standardization has been under way in Optical Wireless Communications Task Group \cite{Trang:18}. As the image sensors are able to separate lighting sources spatially, the reliability and the capacity of the OWC can be enhanced by exploiting a 2D square array of transmit LEDs and high frame rate image sensors \cite{TYamazato:14}.

For OOK systems, the transmitter employs spatial modulation techniques for determining binary transmit LED intensity. Message $b$ is encoded using an $L$-by-$L$ matrix $\mathbf{S}_{b}$ that maps a 2D OOK modulation symbol. Decoding the transmitted message relies on the image captured by a $T$-by-$T$ image sensor. This requires the joint optimization of 2D OOK modulation rule and image decoding process over a signal-dependent optical channel.

The feasibility of the ISC systems has been investigated in indoor scenarios \cite{Cahyadi:16} and outdoor vehicular communication applications \cite{TYamazato:14}. However, the control of the average intensity for the LED arrays, such as dimmable transmitter optimization for the VLC, has not been adequately investigated with the target of the SER minimization. Since high-resolution cameras are adopted in \cite{TYamazato:14} and \cite{Cahyadi:16}, a naive image processing technique, which subtracts the current image from the previous one, suffices to detect on and off symbols conveyed by each LED. By contrast, for a practical low-resolution CMOS image sensor, such an approach would not be possible as it suffers from LED irradiance spread and lens blur.

Other difficulties arise due to the randomness of the ISC channel, in particular, the imperfect alignment between the transmit LED array and the receive image sensor incurring random rotations in the received image. To address this issue, \cite{Cahyadi:16} utilized dummy LEDs, which always emit the same OOK intensity pattern, so that the misalignment can be compensated at the receiver via simple image processing. However, this fails to guarantee arbitrary LED intensity control and results in the degraded spectral and energy efficiency. Generally, designing the ISC transceiver which is robust to the random nature of the optical channel is a highly challenging problem, in particular, when perfect channel knowledge is not available.

\subsection{Convolutional Autoencoder}

\begin{figure*}[!htp]
\begin{center}
\includegraphics[width=5.0in]{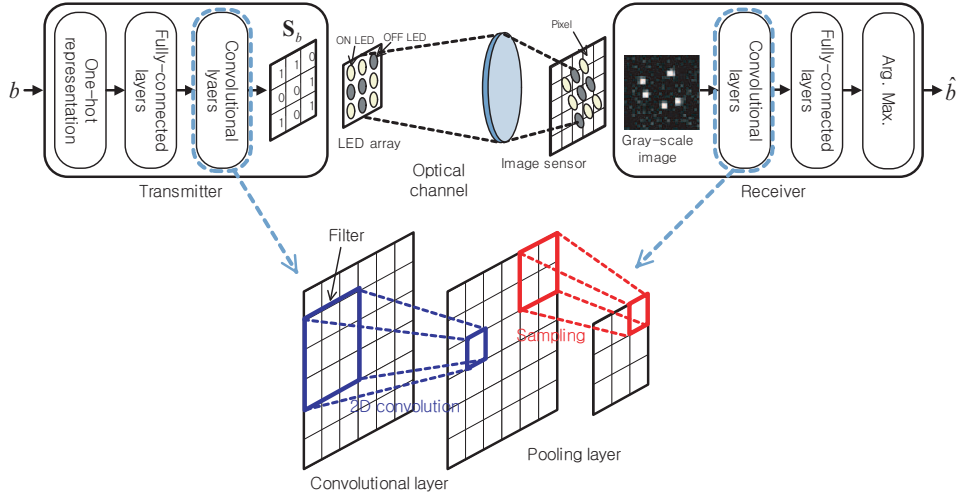}
\end{center}
\caption{Deep learning framework for ISC systems.}
\label{fig:fig3}
\end{figure*}

To overcome the implementation issues in the ISC system, we propose a C-AE structure as illustrated in Fig. \ref{fig:fig3}. In the C-AE, several hidden layers are implemented with convolutional layers, which have proven powerful in handling a 2D image input \cite{LeCun:15}. Unlike one-dimensional (1D) fully-connected layers in Fig. \ref{fig:fig1} where all the elements of an input vector contribute to a hidden layer output, convolutional layers accept matrices as inputs and apply weight matrices only to adjacent elements to produce a 2D output. As depicted in Fig. \ref{fig:fig3}, this can be viewed as a 2D convolution operation which slides a 2D window filter of the same weights over the input matrix. This computation helps extracting spatially-correlated features of the image input such as edges and lines. More complicated features can be learned with the aid of multiple convolution filters having different weights, which provide several 2D output matrices. A pooling layer can be added to a convolutional layer to reduce the output dimension by sampling one element over the predefined 2D region. With the pooling layers, NNs become robust to minor spatial changes in the input image \cite{LeCun:15}. Popular choices for the pooling are the maximum and average operations.

At the encoding network of the C-AE, the message $b\in\{1,\cdots,M\}$ is first mapped to an one-hot vector \cite{OShea:17}, which is a zero vector except for the $b$-th element equal to $1$, and then is processed by several fully-connected layers. To yield a 2D OOK intensity matrix, convolutional layers are adopted to an output vector of the fully-connected layers by reshaping into a matrix. Each element of the output matrix of the encoding network is mapped to the OOK transmit intensity of each LED. Through the optical channel, which contains the signal-dependent noise as well as random image rotation and blur effects, the receiver obtains 2D images capturing the transmit LED array at each data transmission. The decoding network, which includes multiple convolutional layers followed by fully-connected layers, retrieves the transmitted message from the received image.

For implementing the OOK modulation, the parameterized sigmoid activation is adopted at the end of the encoding network with the aid of the annealing-based multi-stage training strategy \cite{HLee:18b}. To control the average intensity of the binary optical signal, the regularization term $\lambda\|\frac{1}{M}\sum_{b=1}^{M}\mathbf{S}_{b}-\mathbf{D}\|_{2}^{2}$, which evaluates the deviation of the average intensity from the $L$-by-$L$ target intensity matrix $\mathbf{D}$, is added to the categorial cross-entropy cost function. Following accurate mathematical ISC channel model in \cite{Ramirez:14}, training samples can be readily generated. To compensate the random rotation effect, the ISC channel is randomly generated with arbitrary rotated LED array coordinates. To be specific, a random rotation angle is applied to training samples so that the C-AE can efficiently extract the features regarding the random rotation by itself. On the other hand, the trained C-AE does not require the rotation angle in the testing step. Thus, the proposed C-AE transceiver can be implemented in a practical scenario where channel state information (CSI) is not available in advance.

\subsection{Implementation Details}

\begin{table}[h!]
\centering
\caption{Proposed C-AE Structure.}
%\subtable[]{
\centering
\begin{tabular}{l|l|l}
\hline
\hline
\multicolumn{3}{c}{Encoding network} \\
\hline
\multicolumn{1}{c|}{Layer}                      & \multicolumn{1}{c|}{Activation} & \multicolumn{1}{c}{\begin{tabular}[c]{@{}c@{}}Output\\ dimension\end{tabular}} \\ \hline
Fully-connected                                 & ReLU                            & $M$-by-$1$                           \\
Fully-connected                                 & ReLU                            & $16L^2$-by-$1$                       \\
Convolutional ($M$ filters, $3$-by-$3$)  & ReLU                            & $4L$-by-$4L$                         \\
Max-pooling ($2$-by-$2$)                        & -                               & $2L$-by-$2L$                         \\
Convolutional ($2M$ filters, $3$-by-$3$) & ReLU                            & $2L$-by-$2L$                         \\
Max-pooling ($2$-by-$2$)                        & -                               & $L$-by-$L$                           \\
Convolutional ($1$ filter, $3$-by-$3$)   & Parameterized sigmoid           & $L$-by-$L$ \\ \hline
\hline
\multicolumn{3}{c}{Decoding network} \\
\hline
\multicolumn{1}{c|}{Layer}                      & \multicolumn{1}{c|}{Activation} & \multicolumn{1}{c}{\begin{tabular}[c]{@{}c@{}}Output\\ dimension\end{tabular}} \\ \hline
Convolutional ($2M$ filters, $5$-by-$5$) & ReLU                            & $T$-by-$T$                           \\
Max-pooling ($2$-by-$2$)                        & -                               & $\frac{T}{2}$-by-$\frac{T}{2}$       \\
Convolutional ($M$ filters, $3$-by-$3$)  & ReLU                            & $\frac{T}{2}$-by-$\frac{T}{2}$       \\
Max-pooling ($2$-by-$2$)                        & -                               & $\frac{T}{4}$-by-$\frac{T}{4}$       \\
Fully-connected                                 & ReLU                            & $M$-by-$1$                           \\
Fully-connected                                 & Softmax                         & $M$-by-$1$ \\ \hline \hline
\end{tabular}
\label{tab:tab1}
\end{table}

A square LED array of size $5$-by-$5$ is considered with a white LED RL5-W4575 \cite{Ramirez:14}. The inter-LED distance is fixed as $1.5\ \text{cm}$, and the distance between the transmitter and the receiver is given by $5\ \text{m}$. The receive image sensor adopts a GigE monochrome 1/4 inch Sony CCD \cite{Ramirez:14} with the resolution of $28$-by-$28$ pixels, each of which has a square shape of size $5.6\ \mu\text{m}$-by-$5.6\ \mu\text{m}$. The lens focal length and the fnumber are set to $3.5\ \text{mm}$ and $1.4$, respectively. The perfect synchronization is assumed between the transmitter and the receiver. The spectral efficiency of $6$ bits per channel use is assumed with $M=64$ messages and the shot noise scaling factor is equal to $\psi^{2}=5$.

The proposed C-AE structure is illustrated in Table \ref{tab:tab1}. At each layer, the batch normalization layer is added for efficient training \cite{Ioffe:15}. Total $10^{6}$ samples are employed for the training, and another $10^{6}$ randomly generated samples are used for the validation step for finding the regularization parameter $\lambda$ which achieves a good tradeoff between the validation SER performance and the dimming feasibility. The test performance of the trained C-AE is evaluated with $10^{9}$ samples. The Adam algorithm \cite{Kingma:15} with the learning rate $0.001$ is utilized for the training of total $7$ stages. To capture the signal-dependent property, two different signal-to-noise ratio (SNR) values, $\text{SNR}_{\text{low}}=10\ \text{dB}$ and $\text{SNR}_{\text{high}}=14\ \text{dB}$, are considered in the training step, each of which is employed in the testing for the low SNR and the high SNR regimes, respectively. A C-AE training step is an offline process. Once trained, real-time operations of the encoding network are carried out in a look-up table that maps a message to the corresponding OOK-modulated symbol. The calculation of the decoding layer is realized by linear algebraic operations of the complexity $\mathcal{O}(T^{2}M^{2})$, which is comparable to maximum-likelihood (ML) detection given by $\mathcal{O}(T^{2}ML^{2})$.

\subsection{Numerical Results}

\begin{figure}
\begin{center}
\includegraphics[width=3.5in]{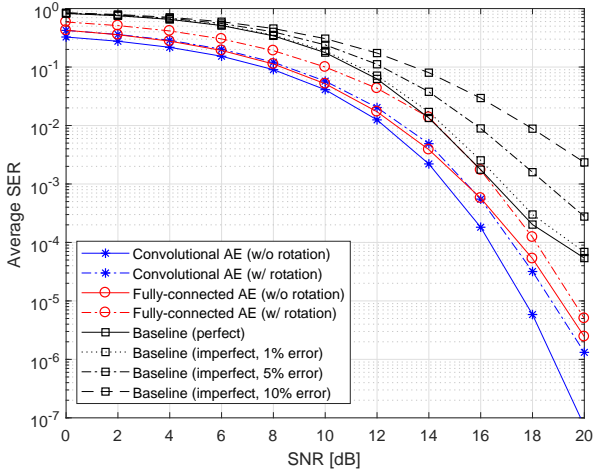}
\end{center}
\caption{Average SER performance as a function of SNR.}
\label{fig:fig4}
\end{figure}

Numerical results for the trained C-AE is presented by evaluating the average SER performance over the testing set comprised with unseen rotation angles and noise. Figure \ref{fig:fig4} plots the average SER performance of the proposed C-AE transceiver as a function of the SNR. All the elements of the target intensity matrix $\mathbf{D}$ are set to $\frac{20}{M}$. Two different scenarios are adopted for the C-AE. First, the C-AE is trained and tested without the rotation to provide reference performance. Second, an arbitrary rotation angle, which is uniformly distributed over $[-30^{\circ},30^{\circ}]$, is applied both in the training and the testing steps of the C-AE. For comparison, the following reference approaches are considered.
\begin{itemize}
\item \textit{Fully-connected AE (F-AE):} The AE only with fully-connected layers is employed with the same number of layers and dimensions as the proposed C-AE structure.
\item \textit{Baseline:} The transmitter utilizes randomly generated OOK satisfying the target average intensity. Then, the ML decoding is applied to the receiver.
\end{itemize}
It is noted that blind detection is possible both for the C-AE and the F-AE without the CSI at the receiver. In contrast, the baseline technique relies on perfect CSI for ML decoding at the receiver. Thus, for fair comparison, the performance of the baseline method is also evaluated for imperfect CSI cases with different level of channel estimation errors. Figure~\ref{fig:fig4} shows that the proposed C-AE outperforms the baseline scheme. The C-AE learns an efficient encoding-decoding rule by observing numerous ISC channels during training, whereas the transmitter and the receiver in the baseline approach are developed separately for a given CSI. It is interesting to see that the C-AE performs better than the F-AE for all SNR range. This implies that 2D convolution operations at the encoding and the decoding of the proposed C-AE are powerful for learning 2D OOK spatial modulation rules as well as image decoding strategies for the ISC systems. Also, the C-AE trained with the random rotation is shown to be robust to the random ISC channel effects, since it provides substantial SER gains over the baseline method with perfect CSI.

\section{Conclusions and Future Works}
This article has introduced DL-based design directions for OWC systems to overcome implementation difficulties stemmed from non-trivial lighting constraints and impairment of optical channels. Recent DL approaches for OWC transceiver optimization have been reviewed and their technical contributions have been discussed. Also, the C-AE framework has been proposed for designing ISC systems where a receiver with an image sensor captures the image of a transmit LED array for decoding. Numerical results have confirmed that the proposed C-AE provides a substantial performance gain over the baseline approaches even without the CSI. Some future research directions are summarized in the following.

\subsection{Dimming-Aware Neural Network Construction}
Current AE methods for VLC design have focused on satisfying a specific dimming constraint, resulting in high computations for training multiple AEs for all possible dimming values. For practical VLC with arbitrary dimming requirement, dimming-aware AE structures which have adaptive dimming control abilities should be investigated. One possible approach is to accept dimming target as an input feature of an NN. With numerous dimming samples, the trained network will be able to support arbitrary dimming requirements via a single training process.

\subsection{Extension to Screen Modulation Systems}
A screen modulation technique \cite{Trang:18}, which conveys a message masked by screen images, is an interesting future research topic. In this case, NNs are trained to modulate color, shape, or intensity of pixels while producing the target image via the RGB intensity control of the screen. Furthermore, a quantization layer with multiple quantization levels, which can be regarded as an extension of the binarization technique for the OOK modulation, is essential to train the NNs for selecting a proper symbol among multiple modulation candidates.

\subsection{Training With Real Measurement Samples and Field Experiments}
Existing DL approaches for OWC stem from mathematical channel models. It is questionable whether trained networks work well in a real-world optical environment that includes LED non-linearity and PD imperfection. Thus, it is necessary to train NNs with the set of measurement samples and validate its performance with field experiments. Recent generative learning techniques such as generative adversarial networks could be exploited to produce high quality of artificial OWC channels based on a small number of measurement samples. The NN is then further trained over these generative samples and its viability can be verified through field experiments.

\subsection{Robust DL Techniques for Communication Systems}
Performance of current wireless networks highly relies on perfect knowledge of modulation and coding schemes, CSI, resource scheduling information, and so on. Acquiring such information would be significantly difficult in 5G systems with a massive number of entities such as large-scale antenna array systems and internet-of-things networks. In these scenarios, developing robust transceivers with imperfect or insufficient prior knowledge is crucial. However, it is not straightforward to handle this issue by existing signal processing methods. Motivated by the results of the proposed ISC systems, the AE technique can be extended to design various wireless systems without the CSI. It would be an interesting future work to investigate a DL framework for developing robust wireless networks where no prior information is available.

\bibliographystyle{IEEEtran}

% Generated by IEEEtran.bst, version: 1.14 (2015/08/26)
\begin{thebibliography}{10}
\providecommand{\url}[1]{#1}
\csname url@samestyle\endcsname
\providecommand{\newblock}{\relax}
\providecommand{\bibinfo}[2]{#2}
\providecommand{\BIBentrySTDinterwordspacing}{\spaceskip=0pt\relax}
\providecommand{\BIBentryALTinterwordstretchfactor}{4}
\providecommand{\BIBentryALTinterwordspacing}{\spaceskip=\fontdimen2\font plus
\BIBentryALTinterwordstretchfactor\fontdimen3\font minus
  \fontdimen4\font\relax}
\providecommand{\BIBforeignlanguage}[2]{{%
\expandafter\ifx\csname l@#1\endcsname\relax
\typeout{** WARNING: IEEEtran.bst: No hyphenation pattern has been}%
\typeout{** loaded for the language `#1'. Using the pattern for}%
\typeout{** the default language instead.}%
\else
\language=\csname l@#1\endcsname
\fi
#2}}
\providecommand{\BIBdecl}{\relax}
\BIBdecl

\bibitem{Khalighi:14}
M.~A. Khalighi and M.~Uysal, ``{Survey on free space optical communication: a
  communication theory perspective},'' \emph{IEEE Commun. Surveys Tuts.},
  vol.~16, no.~4, pp. 2231--2258, 4th Quart., 2014.

\bibitem{SZhao:17}
S.~Zhao and X.~Ma, ``{A spectral-efficient transmission scheme for dimmable
  visible light communication systems},'' \emph{J. Lightw. Technol.}, vol.~35,
  no.~17, pp. 3801--3809, Sept. 2017.

\bibitem{XLiang:17}
X.~Liang, M.~Yuan, J.~Wang, Z.~Ding, M.~Jiang, and C.~Zhao, ``{Constellation
  design enhancement for color-shift keying modulation of quadrichromatic LEDs
  in visible light communications},'' \emph{J. Lightw. Tech.}, vol.~35, no.~17,
  pp. 3650--3663, Sept. 2017.

\bibitem{Ostergard:10}
P.~Ostergard, ``{Classification of binary constant weight codes},'' \emph{IEEE
  Trans. Inf. Theory}, vol.~56, no.~8, pp. 3779--3785, Aug. 2010.

\bibitem{YBengio:09}
Y.~Bengio, ``{Learning deep architectures for AI},'' \emph{Foundat. Trends
  Mach. Learn.}, vol.~2, no.~1, pp. 1--127, 2009.

\bibitem{OShea:17}
T.~O'Shea and J.~Hoydis, ``{An introduction to deep learning for the physical
  layer},'' \emph{IEEE Trans. Cog. Commun. Netw.}, vol.~3, no.~4, pp. 563--575,
  Dec. 2017.

\bibitem{LeCun:15}
Y.~LeCun, Y.~Bengio, and G.~Hinton, ``{Deep learning},'' \emph{Nature}, vol.
  521, pp. 436--444, May 2015.

\bibitem{HLee:18}
H.~Lee, I.~Lee, and S.~H. Lee, ``{Deep learning based transceiver design for
  multi-colored VLC systems},'' \emph{Opt. Express}, vol.~26, no.~5, pp.
  6222--6238, Feb. 2018.

\bibitem{HLee:18b}
H.~Lee, I.~Lee, T.~Q.~S. Quek, and S.~H. Lee, ``{Binary signaling design for
  visible light communication: a deep learning framework},'' \emph{Opt.
  Express}, vol.~26, no.~14, pp. 18\,131--18\,142, July 2018.

\bibitem{Trang:18}
T.~Nguyen, A.~Islam, T.~Yamazato, and Y.~M. Jang, ``{Technical issues on IEEE
  802.15.7m image sensor communication standardization},'' \emph{IEEE Commun.
  Mag.}, vol.~56, no.~2, pp. 213--218, Feb. 2018.

\bibitem{TYamazato:14}
T.~Yamazato, I.~Takai, H.~Okada, T.~Fujii, T.~Yendo, S.~Arai, M.~Andoh,
  T.~Harada, K.~Yasutomi, K.~Kagawa, and S.~Kawahito, ``{Image-sensor-based
  visible light communication for automotive applications},'' \emph{IEEE
  Commun. Mag.}, vol.~52, no.~7, pp. 88--97, July 2014.

\bibitem{Cahyadi:16}
W.~A. Cahyadi, Y.~H. Kim, Y.~H. Chung, and C.-J. Ahn, ``{Mobile phone
  camera-based indoor visible light communications with rotation
  compensation},'' \emph{IEEE Photonics J.}, vol.~8, no.~2, Apr. 2016.

\bibitem{Ramirez:14}
J.~Perez-Ramirez and D.~K. Borah, ``{A single-input multiple-output optical
  system for mobile communication: modeling and validation},'' \emph{IEEE
  Photonics Tech. Lett.}, vol.~26, no.~4, pp. 368--371, Feb. 2014.

\bibitem{Ioffe:15}
S.~Ioffe and C.~Szegedy, ``{Batch normalization: accelerating deep network
  training by reducing internal covariance shift},'' \emph{in Proc. Int. Conf.
  Mach. Learn. (ICML)}, pp. 448--456, July 2015.

\bibitem{Kingma:15}
D.~Kingma and J.~Ba, ``{Adam: a method for stochastic optimization},'' \emph{in
  Proc. Int. Conf. Learn. Represent. (ICLR)}, 2015.

\end{thebibliography}

\input{bibliography.filelist}

\begin{IEEEbiographynophoto}
{Hoon Lee} (hoon$\_$lee@sutd.edu.sg) received the B.S. and Ph.D. degrees in electrical engineering from Korea University, Seoul, South Korea, in 2012 and 2017, respectively.
In 2018, he joined the Singapore University of Technology and Design, Singapore, where he is currently a Post-Doctoral Fellow. His current research interests include machine learning and signal processing for wireless communications, such as visible light communications, wireless energy transfer communication systems, and secure wireless networks.
\end{IEEEbiographynophoto}

\begin{IEEEbiographynophoto}
{Sang Hyun Lee}
(sanghyunlee@korea.ac.kr) received his B.S. and M.S. degrees from Korea Advanced Institute of Science and Technology (KAIST) in 1999 and 2001, respectively, and his Ph.D. degree from the University of Texas at Austin in 2011. Since 2017, he has been with the School of Electrical Engineering, Korea University, Seoul, Korea. His research interests include coding, modulation, optimization and their applications to optical wireless communication.
\end{IEEEbiographynophoto}

\begin{IEEEbiographynophoto}
{Tony Q. S. Quek} (tonyquek@stud.edu.sg) received his B.E. and M.E. degrees in electrical and electronics engineering from Tokyo Institute of Technology, respectively. At MIT, he earned his Ph.D. in electrical engineering and computer science. Currently, he is a tenured associate professor with SUTD. He also serves as the Acting Head of the ISTD Pillar and the Deputy Director of the SUTD-ZJU IDEA. He was honored with the 2008 Philip Yeo Prize for Outstanding Achievement in Research, the 2012 IEEE William R. Bennett Prize, the 2015 SUTD Outstanding Education Awards — Excellence in Research, the 2016 IEEE Signal Processing Society Young Author Best Paper Award, the 2017 CTTC Early Achievement Award, the 2017 IEEE ComSoc AP Outstanding Paper Award, and the 2016-2018 Clarivate Analytics Highly Cited Researcher. He is a Distinguished Lecturer of the IEEE Communications Society and a Fellow of the IEEE.
\end{IEEEbiographynophoto}

\begin{IEEEbiographynophoto}
{Inkyu Lee} (inkyu@korea.ac.kr) received the B.S. degree in control and instrumentation engineering from Seoul National University, Seoul, South Korea, in 1990, and the M.S. and Ph.D. degrees in electrical engineering from Stanford University, Stanford, CA, USA, in 1992 and 1995, respectively. Since 2002, he has been with Korea University, Seoul, where he is currently a Professor with the School of Electrical Engineering. His research interests include digital communications, signal processing, and coding techniques applied for next generation wireless systems. He was a recipient of the Korea Engineering Award from National Research Foundation of Korea in 2017. He was the Chief Guest Editor of the IEEE JOURNAL ON SELECTED AREAS IN COMMUNICATIONS Special Issue on 4G Wireless Systems in 2006. He has served as an Associate Editor for the IEEE TRANSACTIONS ON COMMUNICATIONS, the IEEE TRANSACTIONS ON WIRELESS COMMUNICATIONS, and the IEEE WIRELESS COMMUNICATION LETTERS. He is an IEEE Distinguished Lecturer.
\end{IEEEbiographynophoto}

\end{document}